\documentclass[conference]{IEEEtran}
\IEEEoverridecommandlockouts
\usepackage{makecell}
\usepackage{float}
\usepackage{float}
\usepackage{cite}
\usepackage{amsmath,amssymb,amsfonts}
\usepackage{algorithmic}
\usepackage{graphicx}
\usepackage{textcomp}
\usepackage{xcolor}
\def\BibTeX{{\rm B\kern-.05em{\sc i\kern-.025em b}\kern-.08em
    T\kern-.1667em\lower.7ex\hbox{E}\kern-.125emX}}
\begin{document}
\title{COVID-19 Detection using Transfer Learning  with Convolutional Neural Network \\
}
\author{\IEEEauthorblockN{Pramit Dutta\textsuperscript{1}, Tanny Roy\textsuperscript{2} and Nafisa Anjum\textsuperscript{3} }
\IEEEauthorblockA{\textit{Department of Electronics and }
{Telecommunication Engineering} \\
\textit{Chittagong University of}
{Engineering and Technology}\\
Chittagong, Bangladesh \\
pramitduttaanik@gmail.com\textsuperscript{1},
roy109248@gmail.com\textsuperscript{2} and
nafisaanjum94@gmail.com\textsuperscript{3}
}
}

\maketitle
\begin{abstract}
The Novel Coronavirus disease 2019 (COVID-19) is a fatal infectious disease, first recognized in December 2019 in Wuhan, Hubei, China, and has gone on an epidemic situation. Under these circumstances, it became more important to detect COVID-19 in infected people. Nowadays, the testing kits are gradually lessening in number compared to the number of infected population. Under recent prevailing conditions, the diagnosis of lung disease by analyzing chest CT (Computed Tomography) images has become an important tool for both diagnosis and prophecy of COVID-19 patients. In this study, a Transfer learning strategy (CNN) for detecting COVID-19 infection from CT images has been proposed. In the proposed model, a multilayer Convolutional neural network (CNN) with Transfer learning model Inception V3 has been designed. Similar to CNN, it uses convolution and pooling to extract features, but this transfer learning model contains weights of dataset Imagenet. Thus it can detect features very effectively which gives it an upper hand for achieving better accuracy.
\end{abstract}
\section{Introduction}
The COVID-19(SARS-CoV-2) was originally detected in China at the end of December 2019. COVID-19 became an epidemic all over the world [1]. Statiscally on November 9, 2020 , 508.6 million people were reported as COVID positive in about 216 countries, the number of death crossed 1.2 million whereas more than 35.85 million have recovered [2]. This is a fatal respiratory disease, caused by an acute respiratory syndrome of coronavirus or SARS-CoV-2 virus. According to these statistics, the death rate has been increasing at an alarming rate. Now as medical bio-molecular process, RT-PCR (Reverse Transcription Polymeras Chain Reaction) is one of the standard and effective diagnostic method for diagnosis of COVID-19. But this process is not only time-consuming but also not effecient as RT-PCR is not widely accessible in different countries [1]. On the other hand, CT images are preferable for analysis of tissues of lung muscles as X-RAY cannot effectively distinguish soft tissues in muscles [1].
\par In order to make the process of diagnosing the virus more effective, it has become imperative to develop a time effecient method for the process. With the rapid development of Artificial Intelligence, the Convolutional Neural Networks (CNNs) have been justified as highly effective for the diagnosis of many medical image processing. As such, it becomes easier to detect COVID-19 in Computed Tomography (CT) image. In the proposed model, a multi-layer Convolutional neural network(CNN) has been used. Here, DTL(Deep Transfer Learning) trains the image data by pre-trained model ,to specify, the InceptionV3 model which contains weights from Imagenet.
\par The paper has been categorised into sections with section I introducing the work done here. Section II contains the works related to this field and Section III follows this with the mechanism for the working of proposed model. Data-set, classifier model, image augmentation and hyper-parameter tuning are the subsections included in section III. Section IV contains the result obtained from the designed model. The training process, model evaluation as well as comparison data depicting the effectiveness of proposed technique with respect to existing techniques have been mentioned as subsections of section IV. Section V concludes the paper and is followed by section VI that pertains to the future work in this field.
\begin{figure*}[htbp]
\centering
\includegraphics[width=1.0\textwidth,height=0.25\textheight]{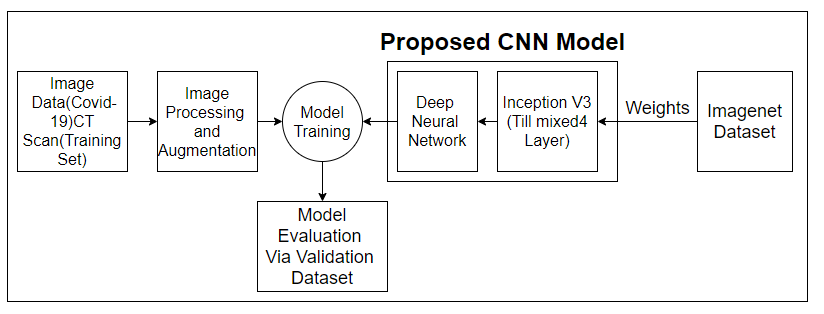}
\caption{Workflow Diagram of the Proposed Model}
\end{figure*}
\section{Related Work}
Throughout recent times, researchers have proposed various methods to detect COVID-19 positive patients and to isolate them as soon as possible. One of these works developed a DenseNet201 based deep transfer learning (DTL) to classify the patients as COVID infected or not i.e. COVID-19 (+) or COVID-19 (-) [1]. Another one of these works has implemented a chest CT image segmentation using a deep convolutional neural network (DCNN), a feature variation block that adjusts the global parameters of the features for segmenting COVID-19 infection was introduced here  [3]. Another paper for detecting COVID-19 has compared various deep learning model performances and developed a model for X-ray, Ultrasound and CT images [4]. Another paper has introduced a deep transfer learning technique and a top-2 smooth loss function with cost sensitive attributes was also utilized for the classification of COVID-19 infected patients [5]. In this paper, an image fusion machine learning model was built by using a deep belief network model [6]. This paper discusses on conventional two-dimensional ultrasound and color Doppler ultrasound to observe the characteristics of lesions in non-critical COVID-19 patients [7]. This paper presents a novel, fully automatic approach for classifying vessels from chest CT images into arteries and veins [8]. Another paper discusses different CNN architectures and evaluates the influence of data-set scale and spatial image context on performance [9].
\section{Methodology}
The system diagram (Fig.1) shows how the whole classification process works. Firstly, the training image was processed and augmented. Then, these images were used to train the neural network built with a combination of transfer learning and DNN model. After this step, the model was evaluated in the validation set. 
\subsection{Dataset}
The COVID-19 CT scan images dataset was collected from kaggle [10]. In this dataset, there were two classes for covid positive and covid negative. For training, there were 279 images for covid positive and also 279 images for covid negative. For validation, the dataset contained 70 images for covid positive and the same amount of images for covid negative. It was 80\% to 20\% split in the training set and validation set respectively.

\subsection{Classifier Model}
In this paper, a deep learning-based neural network model has been proposed in which there are 2 parts. The first part of this network contains a transfer learning model named as Inception V3 and the second part is a customized deep neural network (DNN) layer. This Inception V3 model contains weights of imagenet dataset. The COVID-19 is a newly originated disease. As a result the dataset of covid chest image is not large enough. So, transfer learning model has been used to detect this disease. In this case, Inception V3 has been used which has such filter and convolutional layer that can easily detect features. That gave us a upper hand in the detection of COVID-19. The full structure of this model was not used here rather only a small portion with a DNN was used to extract the feature from that portion.
\subsubsection{Inception V3}
The classification model was developed with the assistance of modified pre-trained Inception-v3 transfer learning model. The Inception-v3 model could factorize the convolution layer which could reduce the number of parameters without even affecting the accuracy. Again, it concatenated the max-pooling and convolutional layer which make feature downsizing more effective. The model had the advantage to extract output from any particular concatenation node. They were named as mixed layer and there are 11 mixed layer in total. In order to make the model more effective for the experiment, the general structure of the model was modified by using only 4 of them as shown in Fig.2. Otherwise, it could cause overfitting due to our small dataset.
\subsubsection{DNN Model}
To extract the output the last few layers of the model were replaced with a DNN which used four customized layers whereas flatten used for transforming mixed4 layer output into one-dimensional array, dense layer had 1024 neurons, the next layer was used for 20\% dropout. Finally, dense\_1 layer with 1 neuron with a sigmoid activation function was used.
\begin{figure*}[htbp]
\centering
\includegraphics[width=1.0\textwidth,height=0.25\textheight]{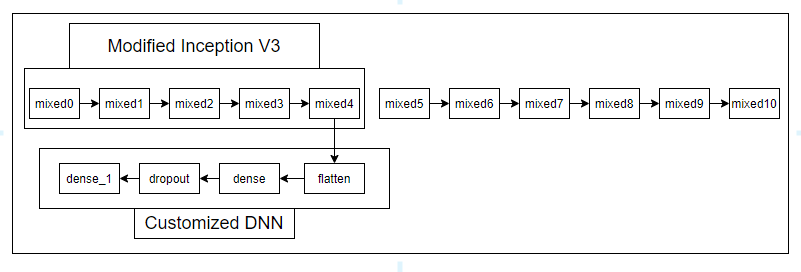}
\caption{Neural Network Model Structure}
\end{figure*}
\subsection{Image Augmentation}
For image detecting purposes it’s an efficient technique to learn more features from image datasets from different dimensions. In this case, the dataset was very small due to insufficient data. So, image augmentation increased the diversity and size of the dataset helping the model to extract unique features. For this purpose, operations like zoom, horizontal flip, shear, shifting were performed on this dataset.
\subsection{Hyperparameter Tuning}
In this model, the RMSprop optimizer was used with a 0.00003 learning rate and batch size of 32. It was a binary classification problem. So, the loss function was binary cross-entropy and metrics were precision and recall.
\section{Result}
\subsection{Training Process}
This model was trained in Google Colab using TensorFlow package. We used  1.66 GB RAM and 1.40 GB GPU. The model was trained for 31 epoches[callback stop] taking about 176 seconds time(5 seconds per epoch).
\subsection{Model Evaluation}
 The model performance was evaluated after every epoch by our validation dataset. In the case of evaluation, four-parameter such as loss, accuracy, precision, and recall were considered.
\begin{table}[H]
\caption{Loss \& Accuracy}
\begin{center}
\begin{tabular}{|c|c|c|c|c|c|}
\hline
\bfseries\makecell{Epoch}&\bfseries\makecell{Training\\ Loss}&\bfseries\makecell{Validation\\ Loss} 
&\bfseries\makecell{Training\\ Accuracy}&\bfseries\makecell{Validation\\ Accuracy} \\
\hline
\makecell{After \\Epoch 1}&1.4701&0.6272&52.51\%&68.57\%\\
\hline
\makecell{After \\Epoch 6}&0.5719&0.7031&67.92\%&58.87\%\\
\hline
\makecell{After \\Epoch 11}&0.4897&0.4438&76.16\%&77.14\%\\
\hline
\makecell{After \\Epoch 16}&0.4370&0.6054&80.65\%&70.00\%\\
\hline
\makecell{After \\Epoch 21}&0.2887&0.4549&86.56\%&83.57\%\\
\hline
\makecell{After \\Epoch 26}&0.3123&0.4597&86.38\%&81.43\%\\
\hline
\makecell{After \\Epoch 31}&0.2073&0.4432&91.40\%&84.29\%\\
\hline
\end{tabular}
\label{tab1}
\end{center}
\end{table}
\par After the training process, loss was 0.2073 and 0.4432 for training and validation following (TABLE I). It can be also be observed from the graph (Fig.3) how training and validation loss are decreasing over epochs. In this case there no major dump(most of the cases between 0.9 to 0.4 for validation set) can be observed during the training process.
\begin{figure}[H]
\includegraphics[width=0.48\textwidth,height=0.2\textheight]{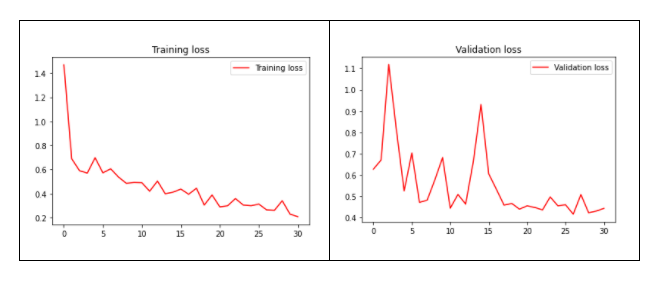}
\caption{Classifiaction Loss vs Epoch}
\end{figure}
On the other hand, after the training process, accuracy was 91.40\% for training and 84.29\% for validation((TABLE I)). Though Fig.4 illustrates some fluctuation during the training, they are small in size (most of the time it fluctuates between 60\% to 75\% for validation set).In a word, our training process was smooth and robust.
\begin{figure}[H]
\includegraphics[width=0.48\textwidth,height=0.20\textheight]{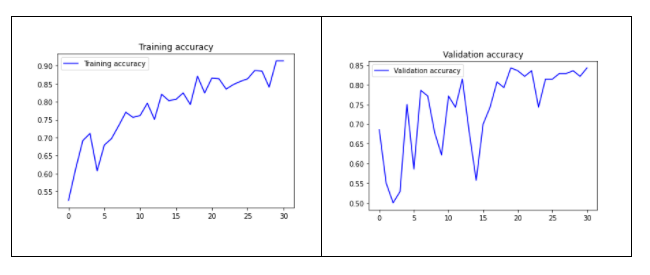}
\caption{Classifiaction Accuracy vs Epoch}
\end{figure}
The model was also evaluated by precision and recall metrics in every epoch and F\textsubscript{1} Score for the final epoch was calculated which also gave a proper evaluation on basis of both precision and recall.
\begin{table*}[t]
\caption{Precision \& Recall}
\begin{center}
\begin{tabular}{|c|c|c|c|c|c|c|c|c|}
\hline
\bfseries\makecell{Factor}&\bfseries\makecell{After\\ Epoch 1}&\bfseries\makecell{After\\ Epoch 6}
&\bfseries\makecell{After\\ Epoch 11}&\bfseries\makecell{After\\ Epoch 16}&\bfseries\makecell{After\\ Epoch 21}&\bfseries\makecell{After\\ Epoch 26}&\bfseries\makecell{After\\ Epoch 31}\\
\hline
\makecell{Training\\Precision}&0.5263&0.6678&0.7664&0.8087&0.8778&0.8638&0.9140 \\
\hline
\makecell{Validation\\Precision}&0.6711&1.0000&0.8519&0.8500&0.7901&0.8548&0.8636\\
\hline
\makecell{Training\\Recall}&0.5018&0.7133&0.7527&0.8029&0.8495&0.8638&0.9068\\
\hline
\makecell{Validation\\Recall}&0.7286&0.1714&0.6571&0.4857&0.9143&0.7571&0.8143\\
\hline
\end{tabular}
\label{tab2}
\end{center}
\end{table*}
\par In the case of precision, it was 0.9140 and 0.8636 for training and validation following (TABLE II). On the other hand training recall was 0.9068 and validation recall was 0.8143 (TABLE II). We know,
\begin{equation}
F\textsubscript{1} Score = 2 \times\left( \frac{Precision \times Recall}{Precision + Recall}\right)
\end{equation}
\par From this equation (1) and result from (TABLE II), 0.9104 was found as training F\textsubscript{1} Score and 0.8382 as validation F\textsubscript{1} Score.
\subsection{Comparison}
MICHAEL. J. HORRY et al [4] discussed on different transfer learning model performance on different types of data. In this proposed model the model performance was evaluated on CT scan data. In the following table, the model performance was compared with other models build for this classification task and used CT scan images for the training and evaluation process.  
\begin{table}[H]
\caption{Comparing Model Performance}
\begin{center}
\begin{tabular}{|c|c|c|}
\hline
\textbf{Model}&\textbf{Accuracy}&\textbf{Training Comment} \\
\hline
VGG-16&79\%&\makecell{Converged.Overfitting\\ evident from 5 epochs.}\\
\hline
VGG-19&78\%&\makecell{Converged. Overfitting\\
evident after 20 epochs.}\\
\hline
Xception&70\%&\makecell{Did not converge. Overfitting\\
evident immediately}\\
\hline
InceptionResnet&63\%&\makecell{Did not converge. Overfitting\\
evident after 1st epoch.}\\
\hline
InceptionV3&71\%&\makecell{Did not converge. Overfitting\\
evident after 2 epochs.}\\
\hline
NasNetLarge&64\%&\makecell{Did not converge. Overfitting\\
evident immediately.}\\
\hline
Densenet121&75\%&\makecell{Converged. Overfitting\\
evident after 8 epochs.}\\
\hline
ResNet50V2&66\%&\makecell{Converged poorly. Overfitting\\
evident immediately.}\\
\hline
Proposed Model&84\%&\makecell{Converged well.Overfitting\\ not evident after 31 epochs.}\\
\hline
\end{tabular}
\label{tab3}
\end{center}
\end{table}
From this table, it can be seen that the previously used VGG-16 model achieved the highest accuracy among all the models working with a CT scan dataset while the inception V3 model had 71\% accuracy. So, our proposed model with 84\% accuracy outperformed all the models proposed in this classification task.
\section{Conclusion}
In this paper, a deep learning concept of transfer learning to detect COVID-19 was proposed. This model shows that computer vision has the power to bring about radical changes in the analysis of radiological images.Hence a time efficient solution can be designed for identifying and isolating infected patients. With a small dataset, the proposed model results in an exceptional result with a validation accuracy of 84\% comparing to 71\% in the InceptionV3 model [3]. This model also outperforms all other models worked with CT Scan data (previously 79\%) (TABLE III). It's because of the combination of InceptionV3 model and the customized deep neural network(DNN) model  that enables this model to outperform all the model that have been proposed based on CT scan images. Moreover, day by day the size of such dataset will increase and so the model will be more accurate and robust.
\section{Future Work}
The future scope of the paper is to test and improve the model performance for other lung diseases as well.

\pagebreak
\end{document}